\documentstyle{mn}

\newcommand{\ie}{{\it i.e.\,}}
\newcommand{\cf}{{\it c.f.\,}}
\newcommand{\etal}{{\it et al.\,}}
\newcommand{\gapprox}{\lower0.8ex\hbox{$\buildrel >\over\sim$}}
\newcommand{\kms}{\mbox{${\,\rm km~s}^{-1}$}\,}
\newcommand{\kev}{\,\mbox{${\,\rm keV}$}\,}
\newcommand{\msun}{\mbox{\,$M_\odot$\,}}
\newcommand{\ergs}{\mbox{${\,\rm erg~s}^{-1}$\,}}
\newcommand{\cmn}{\mbox{${\,\rm cm}^{-2}$\,}}
\newcommand{\cmc}{\mbox{${\,\rm cm}^{-3}$\,}}

\newcommand{\kt}{\mbox{$kT$}}

\newcommand{\rosat}{\mbox{\sl ROSAT}}
\newcommand{\asca}{\mbox{\sl ASCA}}
\newcommand{\einstein}{\mbox{\sl EINSTEIN}}

\title[X-ray Luminous Radio Supernovae in  M82?]
{X-ray Luminous Radio Supernovae in the Center of M82?}

\author[I.\ R. Stevens \etal]{Ian R. Stevens$^{1}$, David K.
Strickland$^{1,2}$ Karen A. Wills$^{3,4,5}$\\ 
$^{1}$ School of Physics and Astronomy, University of Birmingham, 
Edgbaston, Birmingham, B15 2TT, UK\\
$^{2}$ Dept. of Physics and Astronomy, The Johns Hopkins University, 
Baltimore, MD 21218-2686, USA\\
$^{3}$ University of Manchester, Nuffield Radio Astronomy Laboratories,
Jodrell Bank, Macclesfield, Cheshire, SK11 9DL, UK\\
$^{4}$ Department of Physics and Astronomy, University of Manchester,
Oxford Road, Manchester, M13 9PL, UK\\
$^{5}$ Department of Physics, University of Sheffield, Housefield Road,
Sheffield, S3 7RH, UK\\
irs@star.sr.bham.ac.uk; dks@eta.pha.jhu.edu; K.Wills@sheffield.ac.uk}

\date{Accepted .....................; Received .....................; 
in original form .......................}

\begin{document}

\maketitle

\begin{abstract}
We investigate the X-ray emission from the central regions of the
prototypical starburst galaxy M82. Previous observations had shown
a bright central X-ray point source, with suggestions as to its nature
including a low-luminosity AGN or an X-ray binary. A new analysis
of \rosat\ HRI observations find  4 X-ray point sources in
the central kpc of M82 and we identify radio counterparts for the two
brightest X-ray sources. The counterparts are probably young radio
supernovae (SN) and are amongst the most luminous and youthful SN in M82.
Therefore, we suggest that we are seeing X-ray emission from young
supernovae in M82, and in particular the brightest X-ray source is
associated with the radio source 41.95+57.5. We discuss the implications
of these observations for the evolution of  X-ray luminous SN.
\end{abstract}


\section{Introduction}

M82 is the prototypical starburst galaxy, and at X-ray energies is a very
complex source, with a superwind extending out of the plane of the
galaxy for several kpc, and a luminous, point-like central X-ray source
(\cf Watson, Stanger \& Griffiths 1984; Moran \& Lehnert 1997).

In this paper we concentrate on point source emission from the
central regions of M82, rather than the diffuse superwind emission, which
has been discussed elsewhere (\ie Fabbiano 1988; Strickland, 
Ponman \& Stevens 1997).
There has been much speculation as to the nature of the central X-ray
source in M82, which is also variable (Collura \etal 1994). Suggestions
have included a low-luminosity AGN (Tsuru
\etal 1997), inverse Compton emission (Moran \& Lehnert 1997),
an X-ray luminous supernova remnant (Terlevich 1994), a collection of
X-ray binaries (Ptak \etal 1997), a single luminous massive X-ray
binary (Bregman, Schulman \& Tomisaka 1995), or emission 
analogous to the Galactic Ridge emission (Cappi \etal 1999). We note 
that hard X-ray point sources with comparable luminosities 
have been
seen in other starburst galaxies, so that understanding the source in M82
may have more general applicability.

Our goal is to understand the X-ray emission by identifying probable
radio counterparts to X-ray sources in the central regions. Because the
X-ray emission is complex, with diffuse and point-like emission, this
requires careful analysis. We only analyse data from the
\rosat\ HRI instrument, which has the best current spatial resolution ($\sim
5{''}$). We assume a distance to M82 of 3.63\,Mpc (Freedman \etal 1994).

\section{Data and Analysis Procedure}

M82 has been observed by the \rosat\ HRI on three occasions, 
for 53.1ksec, 24.6ksec  and 9.5ksec (see Table \ref{tab:x1}). 
Details of \rosat\ and the HRI
instrument can be found in Briel \etal (1995). To minimise the detector
background we extract images from each HRI observation selecting only
channels 3--8. For all datasets we extracted an image of the entire field,
and used the {\sl ASTERIX} Point Source Searching ({\sl PSS}) package to detect
sources away from the bright central regions. We cross-correlated the
detected X-ray sources with the HST Guide Star Catalogue and the Digitized
Sky Survey. The results were inconclusive, but did not suggest
major pointing errors. The \rosat\ $1\sigma$ pointing
uncertainty is $6{''}$ (Briel \etal 1995, see Fig.~\ref{fig:x1}). The
pointing errors associated with the centroid fitting with the {\sl PSS}
package are typically smaller (Table~\ref{tab:x2}), and
we adopt a $6{''}$ error circle for the positions of the X-ray
point-sources. The impact of the pointing uncertainty is discussed in
Section~3. We also cross-correlated the point sources for the three
X-ray observations, to align the X-ray images. Next, we extracted high
resolution X-ray images of the central kpc of M82, with $2{''}$ pixels,
for detailed source searching in the central regions.

It is vital to accurately account for the extended superwind emission
in M82 in order to reliably identify point sources. To do so we adopted
the following iterative procedure. For our initial background model we
used the non-background subtracted image for each observation, smoothed
with a Gaussian ($20{''}$ FWHM). We source searched this image to
identify the first set of point sources (with a {\sl PSS} $\sigma\geq 4$). For
the second iteration we improved the background model by removing any
sources detected in the first iteration (out to a radius of $7{''}$),
interpolated over the removed source regions and smoothed again. We then
source searched again with this improved background model. We repeated
this process until we had a stable number of sources. The point sources
detected by this procedure are discussed below.

\begin{table}
\caption{\rosat\ HRI observations of M82 used in this paper,
listed in descending order of exposure time}
\begin{tabular}{cccc}
Obs. & Date & Exposure & P.I.  \\
No.  &      & Time &  \\
1 & 1995 Apr. 14 -- May 13 & 53.1\,ksec & P. Serlemitsos\\
2 & 1991 Mar. 25 -- May 04 & 24.6\,ksec & J. Bregman\\
3 & 1992 Oct. 20 -- Oct. 26 & 9.5\,ksec & J. Bregman\\
\end{tabular}
\label{tab:x1}
\end{table}

\section{Results}

\begin{figure}
\vspace*{8cm}
\includegraphics{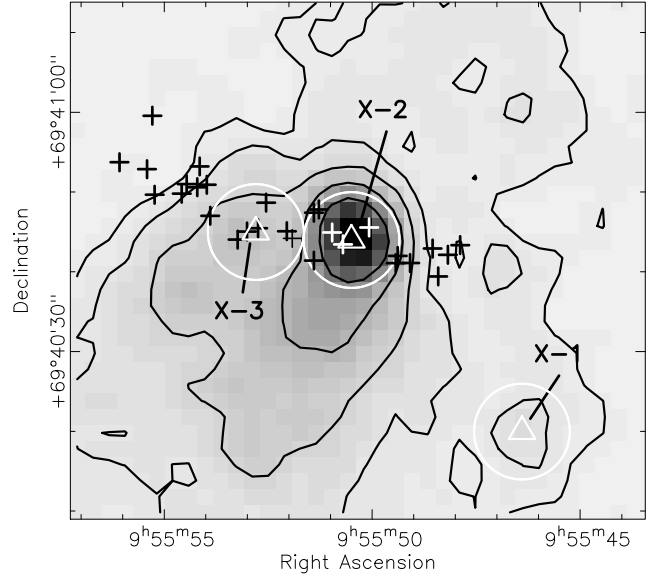}
\caption{The \rosat\ HRI image of the central regions of M82, in the
$0.2-2.0\kev$ waveband. The image has $2{''}$ pixels smoothed with a
Gaussian of FWHM $4{''}$. The contours start at $2.6\times 10^{-2}$
cts$^{-1}$ s$^{-1}$ arcmin$^{-2}$ and increase by a factor 2. The image
is a mosaic of the two shorter observations, as the longest observation
seems to have a problem with blurring (see text for details).
The positions of the X-ray point sources detected are shown with open
triangles and labelled as in Table~2 (source X-4 is off the diagram). The
positions of the radio sources are shown with crosses, and the $6{''}$
radius \rosat\ pointing uncertainties are shown as error circles 
around the point sources.}
\label{fig:x1}
\end{figure}

The X-ray image of the central kpc of M82 reveals a large region of
diffuse emission extending in a roughly NS direction. Combining the
results from all 3 observations we find four X-ray point sources,
shown in Fig.~\ref{fig:x1}, with source details in Table~\ref{tab:x2}.
The brightest point source ({X-2}) is often referred to as the X-ray
nuclear source, but it is not located close to the peak of the $2.2\mu$m
emission. However, source {X-3} {\em is} located close to the
dynamical center of M82 (Wills \etal 1997).

We have cross-correlated the X-ray source positions with those of radio point
sources (Allen \& Kronberg 1998). There are two interesting match-ups.
Source {X-2} is very near to the bright radio source 41.95+57.5,
and source {X-3} is near to 44.01+59.6, a source
previously noted as a possible AGN. The connection between 41.95+57.5 and
the central X-ray source has been noted before (Terlevich 1994) but 
has hitherto not been widely accepted. An X-ray point source
close to the position of 44.01+59.6 was detected by
the \einstein\ HRI (Watson \etal 1984). Here we provide much stronger
evidence for the connection between X-ray point sources and luminous 
SN in M82. Sources {X-1} and {X-4} are located away from the radio sources.

The pointing uncertainty of \rosat\ means that several radio point
sources are within the X-ray error circles of sources {X-2} and {X-3} (see
Fig~\ref{fig:x1}). Consequently, the identification of these X-ray
sources with the respective radio sources is uncertain, and indeed the
X-ray sources may be due to contributions from several of the SN
remnants. Within the $6{''}$ error circles for both X-ray sources there
are several radio point sources (6 for source {X-2}, 7 for
{X-3}). Using the 6cm fluxes from Muxlow \etal (1994) for 
sources within the error circles it is apparent that both 41.95+57.5 and
44.01+59.6 are the brightest sources. For example, the 6cm flux from
41.95+57.5 comprises nearly 90\% of the point source emission from the error
circle around source {X-2}, with the other sources much weaker, 
while 44.01+59.6 contributes nearly 50\% in
the region around {X-3}. In this region the radio source 43.31+59.2 makes
the second largest contribution (and is 40\% as bright as
44.01+59.6). The angular separation of 43.31+59.2 and 44.01+59.6
($4{''}$) makes it impossible to separate them, and it is possible
that 43.31+59.2 does contribute to the X-ray flux from source {X-3}.

While it is by no means universally true that bright X-ray sources always have
bright radio counterparts (as source {X-1} demonstrates) there is often an
association, and the coincidence between the X-ray positions and the two
brightest radio sources seems worth noting.

In the 53.1\,ksec observation the central point source appears extended
(and if a sufficiently high resolution image is binned can actually
appear as a double source). This is probably due to a problem with
the \rosat\ aspect solution for this observation (Morse 1994). For 
this reason we do not include data from this observation in Fig.~\ref{fig:x1}.

\begin{table*}
\caption{The positions of the X-ray point
sources in the central regions of M82, along with the absorption
corrected $L_X$ ($0.1-10.0\kev$ waveband). 
The $1\sigma$ pointing errors associated with the centroid fitting are
also shown.
Note that the $L_X$ for 
sources {X-2} and {X-3} for the 53.1ksec observation (Obs. No. 1) were
calculated assuming a Gaussian PSF ($12{''}$ FWHM) rather than the
\rosat\ HRI PSF (see text for details).}
\begin{tabular}{ccccccc}
Source &\multicolumn{2}{c}{2000 Coordinates}& Centroid & $L_X$ & Obs. & 
Radio   \\
    & RA              & Dec  &  errors & (\ergs) & No. & counterpart \\
    & (09 55)        & ($+69$ 40)  & &         &      & \\
X-1 & 46.4 & 20 & $1.4{''}$ & $1.9\times 10^{39}$ & 2 & -- \medskip\\
X-2 & 50.5 & 44 & $0.3{''}$ & $3.6\times 10^{40}$ & 1 & 41.95+57.5\\
    &      &    & $0.3{''}$ & $4.1\times 10^{40}$ & 2 &\\
    &      &    & $0.7{''}$ & $4.0\times 10^{40}$ & 3 &\medskip \\
X-3 & 52.8 & 45 & $0.5{''}$ & $5.6\times 10^{39}$ & 1 & 44.01+59.6\\
    &      &    & $1.0{''}$ & $6.1\times 10^{39}$ & 2 &\medskip\\
X-4 & 61.7 & 71 & $1.3{''}$ & $1.5\times 10^{39}$ & 2 & -- \\
\end{tabular}
\label{tab:x2}
\end{table*}

\subsection{Radio Counterparts}

41.95+57.5 is the brightest radio point source in M82, and is believed to
be a young SNR. It is sufficiently radio luminous to qualify as a 
\lq radio supernova\rq\ (RSN, a type II supernova exploding
in a dense environment - Van Dyk \etal 1993). 41.95+57.5 is a young
shell-type SN, with a diameter of $0.021{''}$ 
(0.4\,pc; Muxlow \etal 1994), corresponding to a characteristic age of
only 70 years (assuming a constant expansion velocity of $5000\kms$). We note
that this age will be in error if we are dealing with a strongly
decelerating blast-wave in a dense medium. 

44.01+59.6 was considered a possible AGN candidate on 
account of its peculiar radio spectra (Wills \etal 1997), which makes
its detection as an X-ray source of particular note. 44.01+59.6 was
thought to have a positive radio spectral index, as inferred for advection
dominated accretion flows (ADAFs) onto massive black-holes (Yi \& Boughn
1998). Recent evidence suggests that 44.01+59.6 is a young RSN, being radio
luminous but showing a pronounced low frequency cut-off (Allen \&
Kronberg 1998). However, Wills \etal (1999a) suggest 
that 44.01+59.6 may still contain an AGN. 
The radio diameter is $0.05{''}$ (0.9\,pc; Muxlow \etal 1994),
corresponding to a characteristic age of 170 years (assuming an expansion
velocity of $5000\kms$). Both X-ray sources are associated
with super star-clusters (O'Connell \etal 1995), to be
expected as the progenitors were most likely massive stars.

\subsection{X-ray Luminosities}

The HRI instrument does not have any significant spectral capabilities,
so to convert count-rates to fluxes we have to assume a spectral
model. Based on \asca\ results for the central X-ray source we assume a
power-law model with a photon index of $\Gamma=1.7$ and
$N_H=10^{22}\cmn$ (Ptak \etal 1997). The broadband ($0.1-10\kev$) X-ray
luminosities can be estimated using the source count-rates in the soft
($0.2-2.0\kev$) HRI waveband, along with this spectral model.

There are problems with determining the luminosity from the 53.1ksec
observation, as the attitude solution seems to be in error, resulting in
a blurring of the image. Consequently, luminosity estimates
assuming a point source and the HRI PSF results in an underestimate.
Using a larger Gaussian PSF ($12{''}$ FWHM) results
in a substantially higher luminosity - in line with that for the two
other observations. Following the same procedure for the other
observations results in a slightly higher luminosity (as might be
expected), but the increase in luminosity with increasing Gaussian size
is much steeper for the 53.1\,ksec observation. While some variability is
clearly seen from source {X-2} (Ptak \etal 1997) the satellite aspect error
may result in an overestimate of the level of variability.

For the two X-ray point sources with radio counterparts we derive X-ray
luminosities ($0.1-10\kev$, corrected for absorption) of $L_X\sim 4\times
10^{40}\ergs$ for source {X-2} (associated with 41.95+57.5) and $L_X\sim 6
\times 10^{39}\ergs$ for source {X-3} (associated with 44.01+59.6). We shall
discuss these luminosities in the context of X-ray luminous RSN later.

For the other two X-ray point sources we find $L_X \sim 10^{39}\ergs$,
levels attainable by normal massive X-ray binaries (MXRBs), such as SMC\,X-1. 
In the absence of any discernible radio counterparts we conclude that
these sources are likely to be MXRBs.

\section{Discussion}

We have identified 4 X-ray point sources in the central regions of M82
and the probable association between two X-ray sources and luminous RSN in
M82. These two RSN are amongst the youngest and luminous SN in M82, and their
association with the brightest X-ray sources is unlikely to be a
coincidence. Importantly, we can conclude that the bright central X-ray
source in M82 is coincident with a bright RSN. 
The X-ray luminosities of sources {X-2} and {X-3} are consistent with this
model (see below). 

\begin{figure}
\vspace*{7cm}
\includegraphics{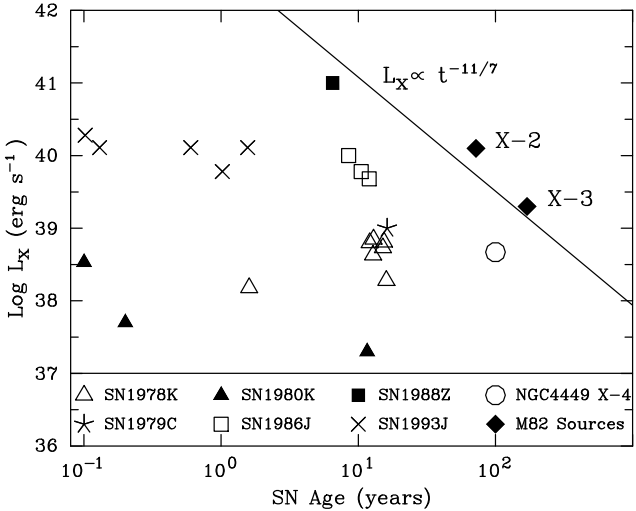}
\caption{The X-ray evolution of X-ray luminous supernova and supernova
remnants. The putative RSN in M82 are shown as solid diamonds sources. A
key is shown to identify the sources. Data for other RSN are from a
range of sources (see text for details). Also shown (solid line) is the
X-ray evolution of a radiative SN remnant in a dense environment
(eqn. 1) for a density of $10^7 \cmc$.}
\label{fig:x2}
\end{figure}

The recent discovery of several X-ray luminous RSN has fueled interest in
their evolution. There exist X-ray observations of young RSN (with ages
$\leq 20$ years; for example SN\,1978K, SN\,1979C, SN\,1980K, SN\,1986J,
SN\,1993J, SN\,1988Z - see Schlegel 1995, Fabian \& Terlevich 1996,
Immler, Pietsch \& Aschenbach 1998), and older SN remnants (ages
$\gapprox 10^3$ years), but the intervening range, in which the two
candidates in M82 probably fall, is poorly sampled. In
Fig.~\ref{fig:x2} we plot $L_X$ versus age for the two M82 RSN as well as
several other young RSN (Schlegel 1995). We also include another example
of an older X-ray luminous RSN, the X-ray source X-4 in NGC\,4449, with
$L_X\sim 5\times 10^{38}\ergs$, and a dynamical age of $\sim 100$\,years
(Vogler \& Pietsch 1997).

In most models of young SN the reverse shock dominates the X-ray emission
(Chevalier \& Fransson 1994). However, Terlevich \etal (1992) suggested
that the forward shock would dominate for a SN explosion in a dense
environment. These objects have been termed compact SNR (cSNR - Terlevich
1994). Fabian \& Terlevich (1996) applied this model to investigate
SN\,1988Z, and we adopt this model here. We note that Wills \etal (1999b)
suggest that 41.95+57.5 is in a very unusual environment, with \lq\lq
chimney\rq\rq\ like features. Assuming a $10^{51}$~ergs SN explosion in
a constant density medium the broadband X-ray luminosity during the
radiative phase will be (Fabian \& Terlevich 1996)

\begin{equation}
L_X (0.1-10\kev) = 7\times 10^{39} n_7^{-3/7} t_{100}^{-11/7} \ergs\ ,
\end{equation}
\noindent with $t_{100}$ the time in units of 100 years, and $n_7$
the ambient density in units of $10^7\cmc$. In Fig.~\ref{fig:x2} we
illustrate the evolution of such an SN, and see that the luminosities for
the M82 sources are broadly consistent with the future evolution of an
SN\,1988Z like object. For SN\,1988Z Fabian \&
Terlevich (1996) estimated an $L_X=10^{41}\ergs$ assuming no local
absorption. SN\,1986J, another X-ray luminous RSN, has a fitted column of
$\sim 5\times 10^{21}\cmn$ (Houck \etal 1998), with the absorption mostly
coming from dense shells of material associated with the SN. If such a
column were applied to SN\,1988Z this would imply a higher luminosity,
with $L_X\sim 5\times 10^{41}\ergs$.

A comment should be made about the radio properties of SN\,1988Z as
compared to the objects in M82. According to Van Dyk \etal\ (1993), at a
time $\sim 5$ years after the SN explosion, the 6cm flux from SN\,1988Z was
1.7mJy and evolving as $t^\beta$ with $\beta=-1.45$. If SN\,1988Z were
at the distance of M82, the flux at the same epoch would be 1.35Jy, and
and at an epoch $\sim 70$ years after the SN explosion the flux would be
30mJy, sufficiently close to the current observed value for 41.95+57.5 
to add weight to the view that the X-ray sources in M82 are similar in
nature to SN\,1988Z, but somewhat older.
The radio spectral indices of SN\,1988Z and SN\,1986J are similar
to that for 41.95+57.5 and 44.01+59.6 (Van Dyk \etal 1993; Allen \&
Kronberg 1999).

Wills \etal (1997) originally suggested 44.01+59.6 may be an AGN on
account of its radio spectrum. Using the results of Yi \& Boughn (1998),
the 2cm radio and X-ray fluxes imply that if the source {X-3} is an ADAF
then the accreting black-hole has a mass of $\sim 5\times
10^7\msun$. Here we have assumed 44.01+59.6 is a RSN (Allen \& Kronberg
1998, though see Wills \etal 1999a). Interestingly, the hardness of the 
spectrum of source {X-2} is
similar to that expected from an ADAF, and this could pose a problem for
the RSN model. The source spectra can be fitted with $\kt\geq 10$\kev
(for a thermal model) or alternatively a power-law with index $\sim 1.7$
(Moran \& Lehnert 1997). 

Terlevich \etal (1992) predicted that the evolution of shock temperature
for a cSNR will be $\propto t^{-10/7}$. Some RSN, younger than the
sources in M82, have measured X-ray temperatures already lower than that
for source {X-2} (\ie SN\,1978K - Ryder \etal 1993; SN\,1986J - Bregman \&
Pildis 1992, though \asca\ observations of SN\,1986J find $\kt\sim
5-7.5\kev$ - Houck \etal 1998). However, Terlevich \etal (1992)
and Plewa (1995) suggest that the spectrum of a single cSNR will be a
power-law, with a slope similar to that observed for source {X-2}.

Consequently, it is not clear whether the observed spectra of {X-2} is at odds
with that expected. However, we do mention the following possibility;
for older SN remnants the X-ray emission can be dominated by non-thermal
emission (\ie SN\,1006, Koyama \etal 1995). Keohane \etal (1997) report
on two hard X-ray emitting regions in the SN remnant IC\,443, which they
interpret as being due to non-thermal emission (with power-law spectra with
$\Gamma=1.3\pm 0.2$). They find that the non-thermal regions, which
dominate the hard X-ray emission from the remnant, are
located on the edges of the remnant, where the shock-wave is interacting
with dense molecular gas. Keohane \etal (1997) discuss their results in
terms of the SN shock/dense cloud interaction model of Jones
\& Kang (1993), which predicts strong particle acceleration and
non-thermal X-ray emission from such a situation.

The scenario of a
strong SN shock encountering dense gas (either a molecular cloud or wind
material from an earlier evolutionary phase) is exactly what would be
expected in the central regions of M82. Thus a strong
SN shock/dense gas interaction will result in strong non-thermal X-ray
emission. In this situation we might also expect there to be significant
variability, in line with what is seen for source {X-2}. Observations with
{\sl Chandra} and {\sl XMM} will constrain whether the X-ray spectrum of
41.95+57.5 is thermal or non-thermal in nature. 

Hard (and variable) point-like sources have been seen in the centres of
other starbursts (\cf Ptak \etal 1997; Dahlem, Heckman \& Fabbiano 1995).
We speculate that X-ray luminous RSN could account
for these sources, without resorting to a low-luminosity AGN or
MXRBs with extremely massive ($\geq 75\msun$) black-holes.

Ultimately it may prove to be difficult to determine whether the X-ray
emission from the sources is due to an AGN or a RSN on X-ray
grounds. Long-term variability will be important, with a long-term
decline indicative of an RSN, while stochastic variability may favour an
AGN (although short-timescale variability is also predicted in the cSNR
model, Tenorio-Tagle 1994). In fact, from Collura \etal (1994),
source {X-2} was about 50\% fainter in \einstein\ observations made in
1979, as compared to \rosat\ observations made in 1991. Source {X-3} was
also detected by the \einstein\ HRI (Watson \etal 1984).
Assuming the spectral model and distance to M82 adopted here we infer
$L_X\sim 10^{40}\ergs$, somewhat above that measured by
\rosat. The X-ray lightcurves shown in Collura \etal (1994) and Ptak \&
Griffiths (1999) appear stochastic on timescales of days and months and
would seem to favour the AGN model, although it must be stressed that
there are no sufficiently detailed observations of the X-ray variability
of RSN. If indeed the SN explodes in a clumpy environment such stochastic
variability would be expected.

In summary, we find that the two brightest X-ray point sources in the
central regions of M82 are likely associated with two bright radio
objects (41.95+57.5 and 44.01+59.6). While there is some pointing
uncertainty in the \rosat\ observations, the alignment between the
brightest X-ray point sources and the most luminous RSN in M82 is 
unlikely to be a coincidence. Both radio counterparts are believed
to be young RSN with ages $\sim 100$ years. The X-ray emission could
be dominated by non-thermal emission associated with a shock/cloud
interaction, though verification of this will require more observations.
These objects begin to fill an important gap in our understanding of SN
evolution. {\sl Chandra} observations will be key in confirming their nature 
and monitoring their X-ray evolution.

\end{document}